\documentclass[final,1p,times]{elsarticleTC0}

\usepackage{amsmath,amssymb,amsfonts,amscd}
\usepackage{graphicx}

\usepackage{caption}
\usepackage{subcaption}

\usepackage{color,soul}
\begin{document}
\begin{frontmatter}

\title{Asymmetric quantum transport in a double-stranded \\ Kronig-Penney model}

\author
{Taksu Cheon}
\ead{taksu.cheon@kochi-tech.ac.jp}
\corref{cor1}

\author
{Sergey S. Poghosyan}
\ead{sergey.poghosyan@kochi-tech.ac.jp}
\address
{Laboratory of Physics, Kochi University of Technology,
Tosa Yamada, Kochi 782-8502, Japan}
%
\date{}	

\begin{abstract}
We introduce a double-stranded Kronig-Penney model and analyze its transport properties.  The asymmetric fluxes between two strands with suddenly alternating localization patterns are found as the energy is varied.  The zero-size limit of the internal lines connecting two strands is examined using quantum graph vertices with four edges.  We also consider a two-dimensional Kronig-Penney lattice with two types of alternating layers with $\delta$ and $\delta'$ connections, and show that the existence of energy bands in which the quantum flux can flow only in selected directions.
\end{abstract}

\begin{keyword}
solvable quantum mechanics \sep periodic potential \sep exotic vertex
\PACS 03.65.-w, 03.65.Nk, 73.63.Nm
\end{keyword}

\end{frontmatter}
%

\section{Introduction}

The investigation of wave propagation through periodic networks \cite{Avishai, Gratus} has been motivated by problems of free-electron motion on organic molecules and quantum wires. The application of more complicated graph structures was found also in other areas such as quantum chaos, photonic crystals, dynamical systems and nanotechnologies \cite{Kuchment}. Mathematically, quantum graphs arise when one considers a propagation of waves through quasi-one-dimensional systems in the limit cases.
The study of periodic quantum graphs with {\it non-conventional graph vertices} \cite{CS06} has brought a new vista to the problem by revealing exotic particle dynamics, which were not found in models with the commonly used $\delta$-potential type vertices. \cite{Ex95, hiroaki,Ex14}

In this paper, we follow that lead and consider double-stranded periodic quantum graphs in the form of chained rings, with non-conventional four-way junction nodes.  The model can be regarded as the two-lane extension of celebrated Kronig-Penney model, \cite{KP31} which is a prototype of solvable system possessing of the band spectra. 
We show that, with the proper tuning of parameters, this system exhibits the energy bands in which quantum fluxes show the asymmetry between two strands, both in terms of relative strengths and direction.  Moreover, the mode of asymmetry goes through a sudden change when the particle energy is varied and the system moves from one band to another.  The pattern of change of the flux asymmetry is shown to be quite various depending on the choice of system parameters.

In order to check that the flux asymmetry is not a particular feature of the specific model of chained rings, we also consider a ladder type double-stranded model with two straight strands connected by periodic internal lines.  For this model we employ a three-way graph vertex of $\delta$ and $\delta'$ type couplings.  We confirm that the flux asymmetry that varies from one band to the other is observed in this type of model as well.

It is essential that we are able to approximate Kronig-Penney lattices with non-conventional junctions, in terms of physically realizable vertex couplings.  As is well known, the $\delta$-potential vertex can be realised as the short-range limit of any regular potential with scaled potential strength that keeps its volume integral constant.  It has been recently found that the vertices with the most general singular couplings can be approximated by auxiliary graphs, carrying only $\delta$ potentials and a magnetic field \cite{approx-gen}.  We explicitly construct finite ladders with only $\delta$ couplings, thus presenting the experimentally constructible double-stranded periodic quantum graph with two outer strands and many internal lines.

The paper is organised as follows:  In section 2, a double-stranded Kronig-Penney model in the form of chained rings is presented, and its characteristics of flux asymmetry is shown.  In section 3, an alternate extended Kronig-Penney model in the form of a ladder is analyzed. In section 4, the natural extension of the asymmetric quantum ladder to the two-dimensional rectangular lattice with alternating $\delta$, $\delta'$ layers is considered.  Section 5 details the method to realize  both mathematical models with conventional $\delta$ coupling vertices.  The last section, section 6, concludes with a summery.

\section{Double-stranded chain with singular vertices}

\begin{figure}[ht]\center
\includegraphics[width=70mm]{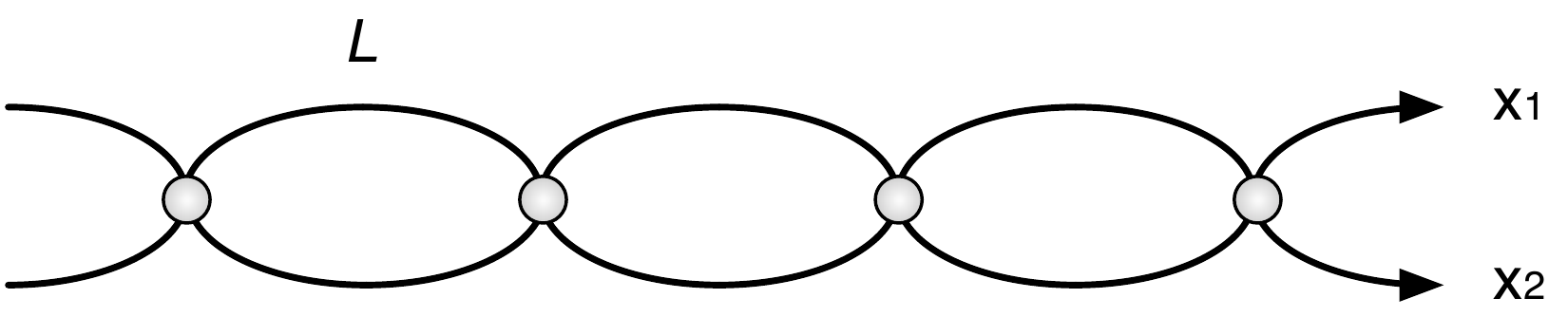}
\caption{Rings of double-stranded Kronig-Penney lattice connected by nodes with singular interactions.}
\label{fig1}
\end{figure}

\begin{figure}[ht]\center
\includegraphics[width=50mm]{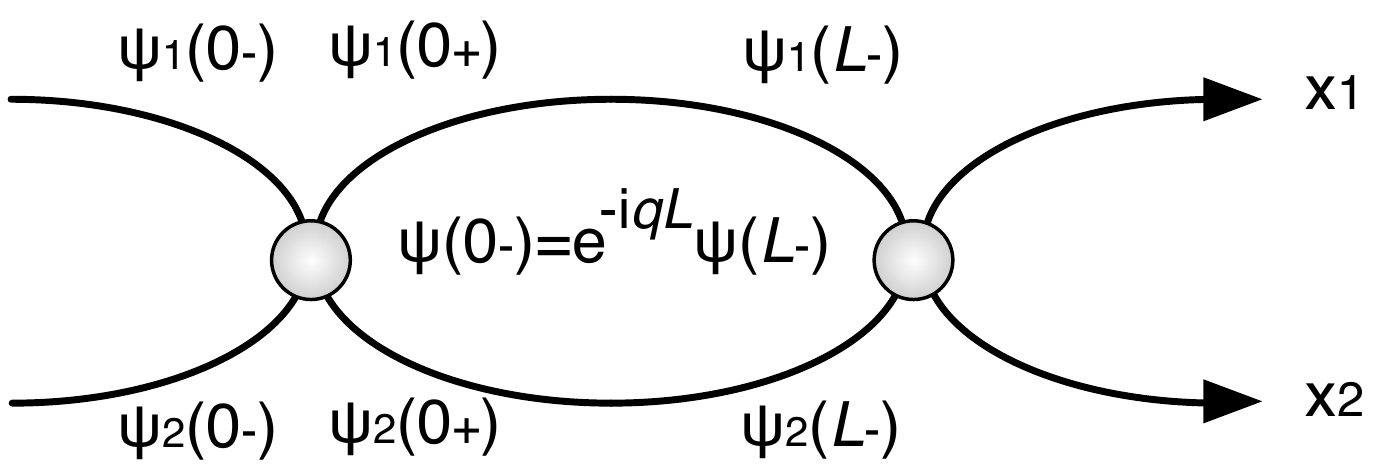}
\caption{Elementary cell of the lattice and quasi-periodicity of wave function.}
\label{fig2}
\end{figure}

We start our investigation of periodic lattice structures containing non-conventional vertex couplings with the simplest case, i.e., the double-stranded Kronig-Penney model with quantum graph vertices (Fig. \ref{fig1}).
We examine the energy eigenstates of a quantum particle residing on this structure.  Due to the Bloch theorem, it is sufficient to consider a single unit segment of the lattice $x_1 \in [0, L)$ and $x_2 \in [0, L)$  as shown in Fig. \ref{fig2}.
The wave function is given in the form
\begin{eqnarray}
\label{solwv1}
&
\psi_1(x_1) = \alpha_1 e^{i k x_1} + \beta_1 e^{-ikx_1} ,
\nonumber \\
&
\psi_2(x_2) = \alpha_2 e^{i k x_2} + \beta_2 e^{-ikx_2} 
\end{eqnarray}
and from the quasi-periodicity condition we have
\begin{eqnarray}
&
\psi_1(0_-) = e^{-i q L} \psi_1(L_-) ,\quad
\psi'_1(0_-) = e^{-i q L} \psi'_1(L_-) ,
\nonumber \\
&
\psi_2(0_-) = e^{-i q L} \psi_2(L_-) ,\quad
\psi'_2(0_-) = e^{-i q L} \psi'_2(L_-) ,
\end{eqnarray}
which yield the relations
\begin{eqnarray}
&
\psi_1(0_-) = e^{-i q L}( e^{i k L}\alpha_1 + e^{-i k L}\beta_1) ,\quad
\psi'_1(0_-) = ik e^{-i q L}( e^{i k L}\alpha_1 - e^{-i k L}\beta_1) ,
\nonumber \\
&
\psi_2(0_-) = e^{-i q L}( e^{i k L}\alpha_2 + e^{-i k L} \beta_2) ,\quad
\psi'_2(0_-) = ik e^{-i q L}( e^{i k L}\alpha_2 - e^{-i k L} \beta_2) .
\label{limfunc}
\end{eqnarray}
The most general connection condition at the graph vertex $x=0$ is given in terms of the vectors 
\begin{eqnarray}
\Psi = \begin{pmatrix} \psi_1(0_-) \\ \psi_1(0_+) \\ \psi_2(0_-) \\ \psi_2(0_+) \end{pmatrix}
\ ,
\Psi'  = \begin{pmatrix}  -\psi'_1(0_-) \\ \psi'_1(0_+) \\ -\psi'_2(0_-) \\ \psi'_2(0_+) \end{pmatrix}
\ ,
\end{eqnarray}
by
\begin{eqnarray}
B \Psi' + A\Psi = 0 ,
\label{AB_boundcond}
\end{eqnarray}
where $A$ and $B$ are complex $4\times 4$ matrices and satisfy the requirements
\begin{equation}\label{KS}
\begin{split}
\bullet \quad & \mathrm{rank}(A|B)=4,\\
\bullet \quad & \text{the matrix $AB^*$ is self-adjoint},
\end{split}
\end{equation}
where $(A|B)$ is a $4\times8$ matrix formed from columns of $A$ and $B$.\cite{KS99}  The choice of matrices $A$ and $B$ is not unique as the equation (\ref{AB_boundcond})
can be multiplied from the left by an arbitrary non-degenerate matrix. 
One convenient way of representing connection conditions (\ref{AB_boundcond}) is transforming them into the \emph{$ST$-form}~\cite{approx-gen}, which is the  block form 
\begin{equation}
\label{ST}
\left(\begin{array}{cc}
I^{(m)} & T \\
0 & 0
\end{array}\right)\Psi'=
\left(\begin{array}{cc}
S & 0 \\
-T^* & I^{(4-m)}
\end{array}\right)\Psi\,,
\end{equation}
where $m\in\{0,1,2,3,4 \}$.  Here, $I^{(m)}$ is the identity matrix $m\times m$, $T$ is a complex matrix of size $m\times (4-m)$ and $S$ is a Hermitian matrix of order $r$. 
%
%
%
%
Since the wave functions at $x=0_+$ are given in terms of coefficients $\alpha_j$ and $\beta_j$ as
\begin{eqnarray}
&
\psi_1(0_+) = \alpha_1 + \beta_1 ,\quad
\psi'_1(0_+) = ik(\alpha_1 - \beta_1) ,
\nonumber \\
&
\psi_2(0_+) = \alpha_2 + \beta_2 ,\quad
\psi'_2(0_+) = ik(\alpha_2 - \beta_2) ,
\end{eqnarray}
the quasi-periodicity conditions of wave function (\ref{limfunc}) can be rewritten in terms of four-vector
\begin{eqnarray}
V = \begin{pmatrix} \alpha_1 \\ \beta_1 \\ \alpha_2 \\ \beta_2 \end{pmatrix},
\end{eqnarray}
as
\begin{eqnarray}
\Psi = e^{-iqL} F V, \quad \Psi' = ik e^{-iqL} G V, 
\end{eqnarray}
where the $4\times4$ matrices $F$ and $G$ are defined by
\begin{eqnarray}
F = 
\begin{pmatrix}
e^{ikL} & e^{-ikL} & 0 & 0 \\
e^{iqL} & e^{iqL} & 0 & 0 \\
0 & 0 & e^{ikL} & e^{-ikL} \\
0 & 0 & e^{iqL} & e^{iqL} \\
\end{pmatrix},
\quad
G =
\begin{pmatrix}
-e^{ikL} & e^{-ikL} & 0 & 0 \\
e^{iqL} & -e^{iqL} & 0 & 0 \\
0 & 0 & -e^{ikL} & e^{-ikL} \\
0 & 0 & e^{iqL} & -e^{iqL} \\
\end{pmatrix}.
\end{eqnarray}
%
In principle, a scaling parameter is required in the connection conditions (\ref{AB_boundcond}) or (\ref{ST}) to compensate the differences in the length dimensions of $\Psi$ and $\Psi'$ \cite{FT00}.  But in practice, it could be dropped since it can be absorbed into the definition of the length unit.
The equation specifying the connection condition $A\Psi + B\Psi' = 0$ now reads
$(A F + ik B G) V = 0$, and a nontrivial solution can be obtained only with the condition
\begin{eqnarray}
\det \left[ {A F + ik B G}\right] = 0 .
\end{eqnarray}
Since the whole family of all possible vertex couplings is too large to handle, we limit
our choice to a subfamily of $m=2$ in the form
\begin{eqnarray}
\label{cc-ddp}
T = 
\begin{pmatrix} a & b\\ a & b  \end{pmatrix} 
\ , \quad
S =
\begin{pmatrix}  c & cd \\ cd & cd^2 \end{pmatrix}.
\end{eqnarray}
For such boundary conditions we obtain
an explicit solution in the form 
\begin{eqnarray}
\label{secul11}
\cos qL= \frac{R\pm\sqrt{R^2+2abkS}}{4abk} ,
\end{eqnarray}
where
\begin{eqnarray}
&&\!\!\!\!\!\!\!\!\!\!\!\!\!\!\!\!\!\!\!\!\!\!\!\!    
R = (a-b)^2 k \cos kL + c [(a^2+b^2)(d-1)^2+d^2+1] \sin kL,
\nonumber \\
&&\!\!\!\!\!\!\!\!\!\!\!\!\!\!\!\!\!\!\!\!\!\!\!\!    
S = (2a^2+2b^2-1)k + (2a^2+2b^2+1) k \cos 2kL - c [(a^2\!+\!b^2)(d-1)^2+d^2+1] \sin 2kL.
\end{eqnarray}
\begin{figure}[ht]\center
\begin{tabular}{cc}
\includegraphics[width=52mm]{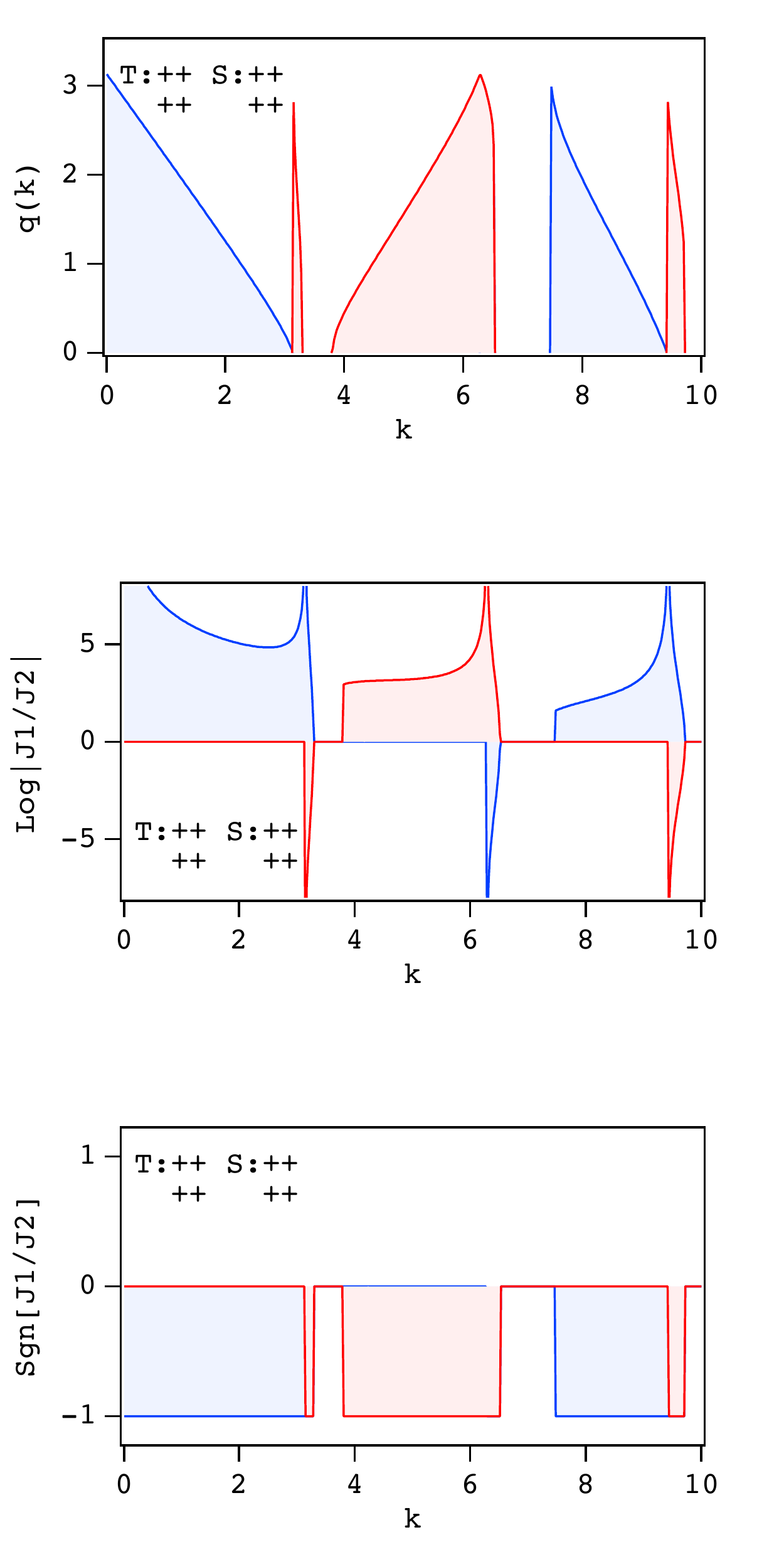}
&
\includegraphics[width=52mm]{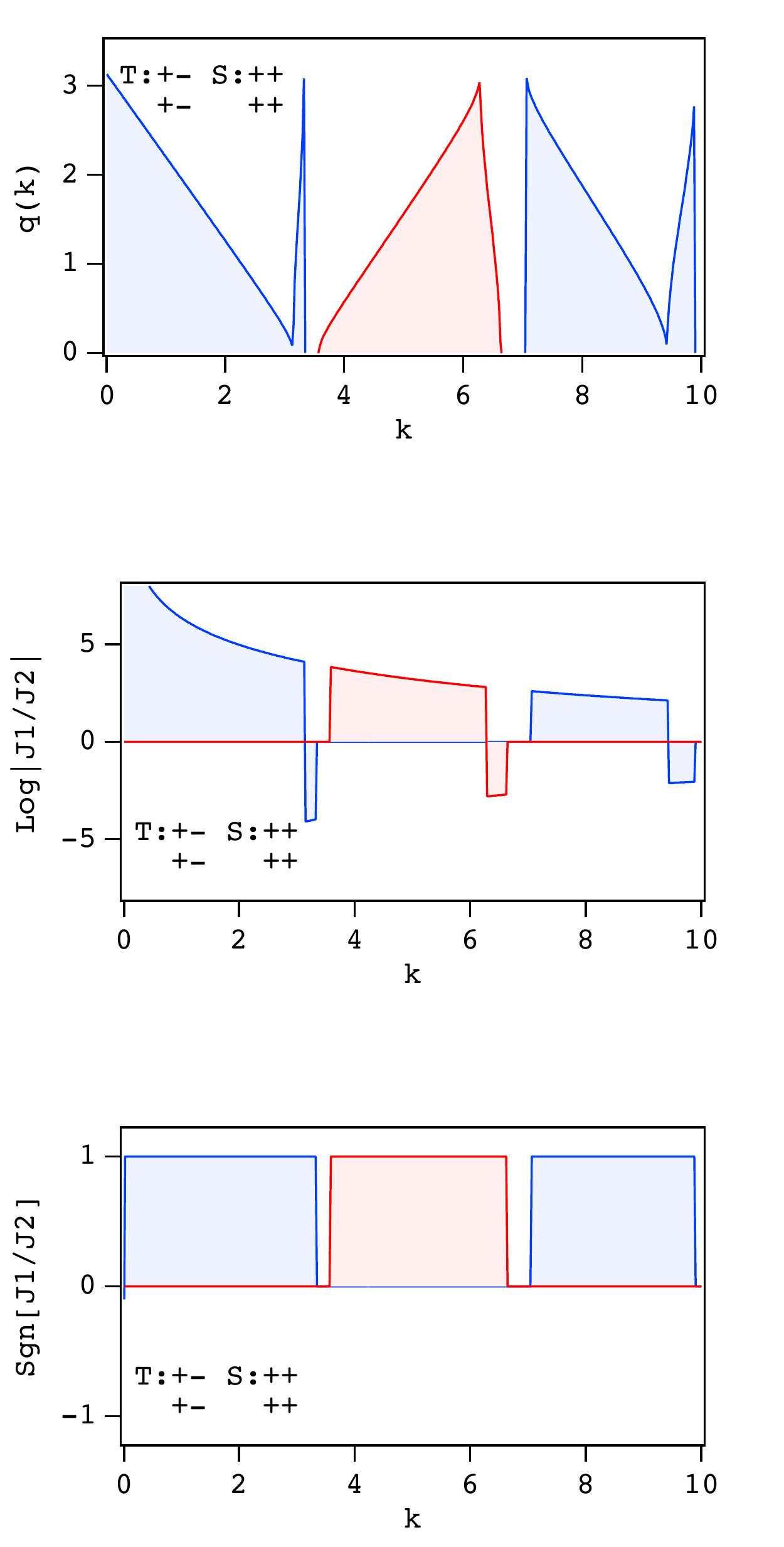} \\
(a)
&
(b)
\end{tabular}
\caption{The band structure (top), the logarithm of the absolute value of flux ratio (middle), and the relative flux direction (bottom) are plotted as the function of the particle energy for the coupling parameter 
(a) $a=b=\frac{1}{3}$, $c=8$, $d=1$, and (b) $a=\frac{1}{3}$, $b=-\frac{1}{3}$, $c=8$, $d=1$. The red and blue lines correspond to the two solutions of secular equation (\ref{secul11}), and the shaded areas indicate energy bands of the spectrum. The plus and minus signs in the inlet caption refer to the structure of  matrices $T$ and $S$ in (\ref{cc-ddp}).}
\label{fig3}
\end{figure}
%
\begin{figure}[ht]\center
\begin{tabular}{cc}
\includegraphics[width=52mm]{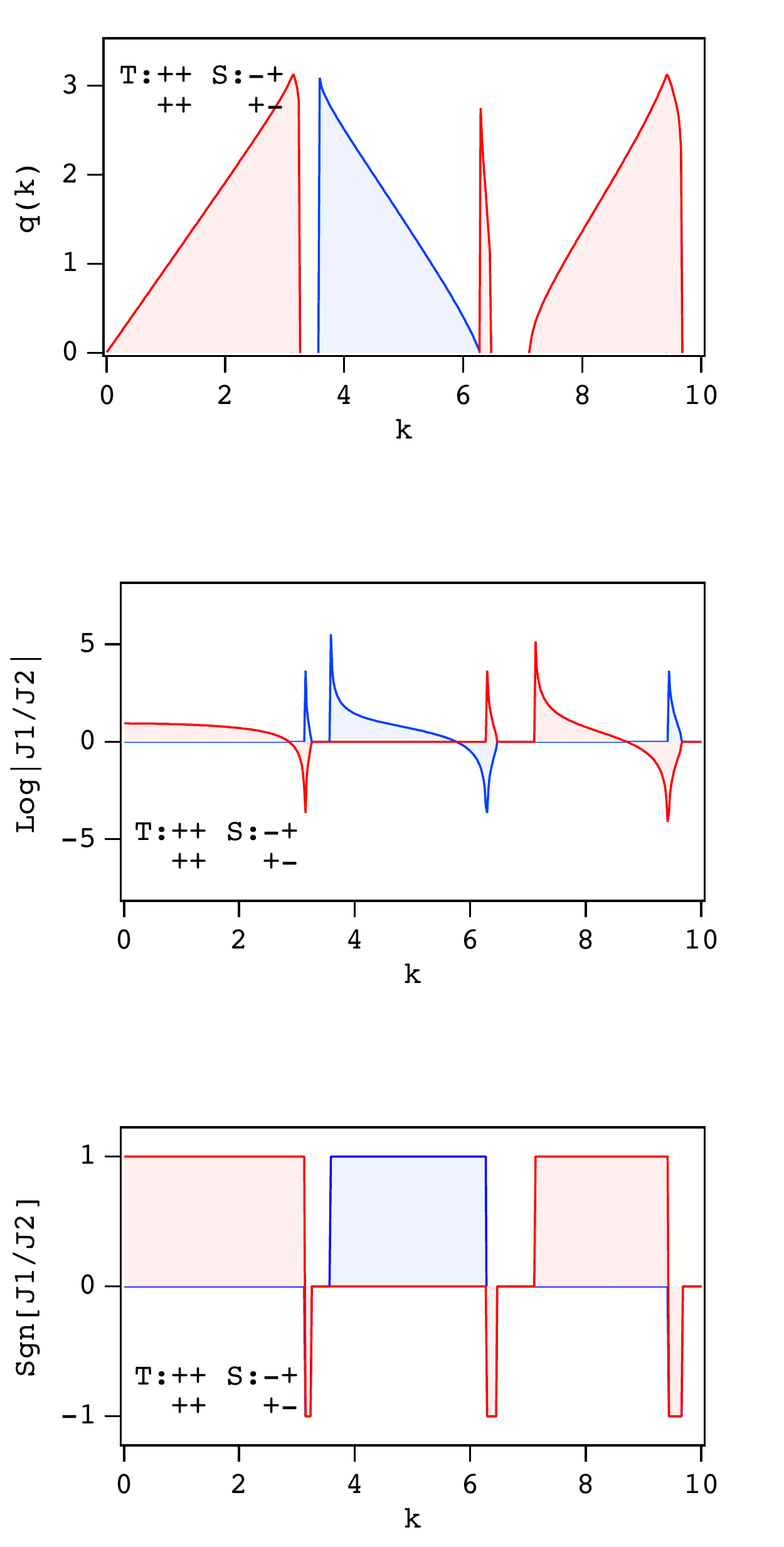}
&
\includegraphics[width=52mm]{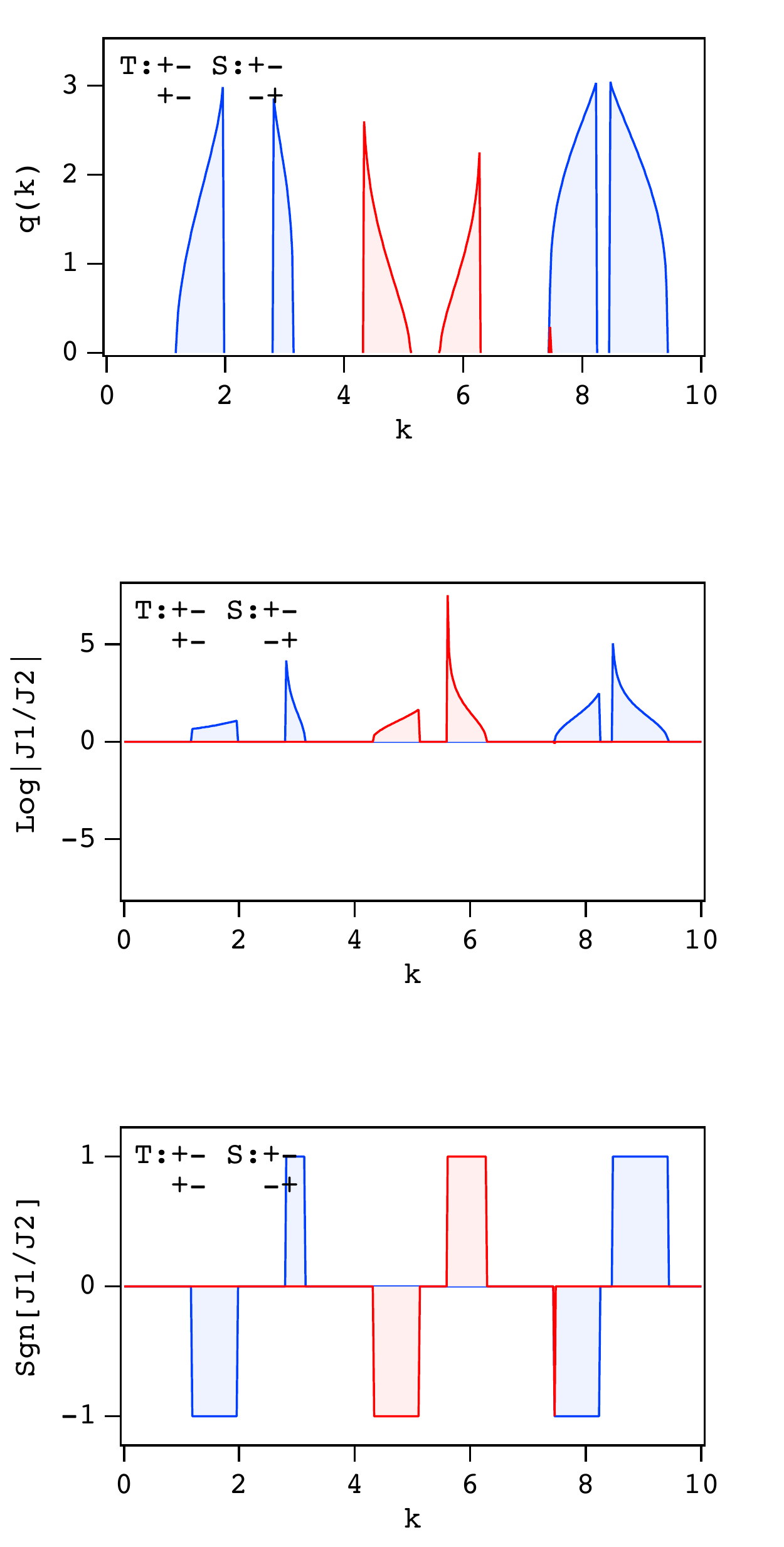} \\
(a)
&
(b)
\end{tabular}
\caption{The band structure (top), the logarithm of the flux ratio (middle), and the relative flux direction (bottom) are plotted as the function of the particle energy for the coupling parameter 
(a) 
$a=b=\frac{1}{3}$, $c=8$, $d=-1$, and (b) $a=\frac{1}{3}$, $b=-\frac{1}{3}$, $c=8$, $d=-1$. 
The red and blue lines correspond to the two solutions of secular equation (\ref{secul11}), and the shaded areas indicate energy bands of the spectrum. the plus and minus signs in the inlet caption refer to the structure of  matrices $T$ and $S$ in (\ref{cc-ddp}).}
\label{fig4}
\end{figure}
Note that there are two sets of bands, which may overlap, each corresponding to plus and minus of the composite signs.
For each band, we calculate
the quantum flux of the particle going through the top and bottom strands, which 
are obtained from the formula $J_j=\frac{\hbar}{2mi}\left(\psi_j^*\psi'_j-\psi_j'^*\psi_j\right)$ $(j=1,2)$ with wave functions (\ref{solwv1}).
We have
\begin{eqnarray}
J_1=\frac{k\hbar}{m}\left(|\alpha_1|^2-|\beta_1|^2\right),\nonumber\\
J_2=\frac{k\hbar}{m}\left(|\alpha_2|^2-|\beta_2|^2\right).
\label{currents12e}
\end{eqnarray}
Notice that from the Bloch theorem, it follows that the currents (\ref{currents12e}) are conserved along the chain.
As we did not fix the boundary conditions of the whole lattice, the wave function is not normalizable and the absolute values of $J_i$ carry no physical meaning.
The ratio between them is physical, on the other hand, and it is the exact quantity of our primary interest:  Its absolute value is the measure of the asymmetry of current flows between two strands, and its sign determines their relative directions.

In Fig. \ref{fig3} and Fig. \ref{fig4} there are shown four sets of numerical examples obtained with combinations $b = \pm a$ and $d = \pm 1$.  The values of other parameters are set to be $a=\frac{1}{3}$ and $c = 8$.  For each set of parameters, we display the energy bands, the logarithm of $|J_1/J_2|$, and the relative sign ${\rm sgn}[J_1/J_2]$, each as functions of particle energy.

The asymmetric quantum transport, with respect to the two strands, is evident.  In each of the alternating regions within a band, the transport is supported mainly in one of the two strands.  With different choices of system parameters, various patterns of asymmetry variations 
are observed as we move from one energy band to the next one.

Particularly intriguing examples of alternating flux asymmetry 
are found in the case $b = -a$, depicted 
in Figs. \ref{fig3} (b) and \ref{fig4} (b);  With the choice $d>0$, we 
find that the localization of flux strength changes
from one strand to another in successive bands, while the relative direction stays
positive all the way, as seen from Fig. \ref{fig3} (b). In the case $d<0$, we observe
that the relative flux direction change
in successive bands, while the relative strength remains
one-sided, as can be noticed from Fig. \ref{fig4} (b).

\section{Asymmetric Kronig-Penny ladder}

In this section we will give a realization of a double-stranded Kronig-Penney system by asymmetric periodic ladder with internal lines. We will obtain the chained ring model of the previous section as a zero-size limit case of internal lines.

\begin{figure}[ht]\center
\includegraphics[width=70mm]{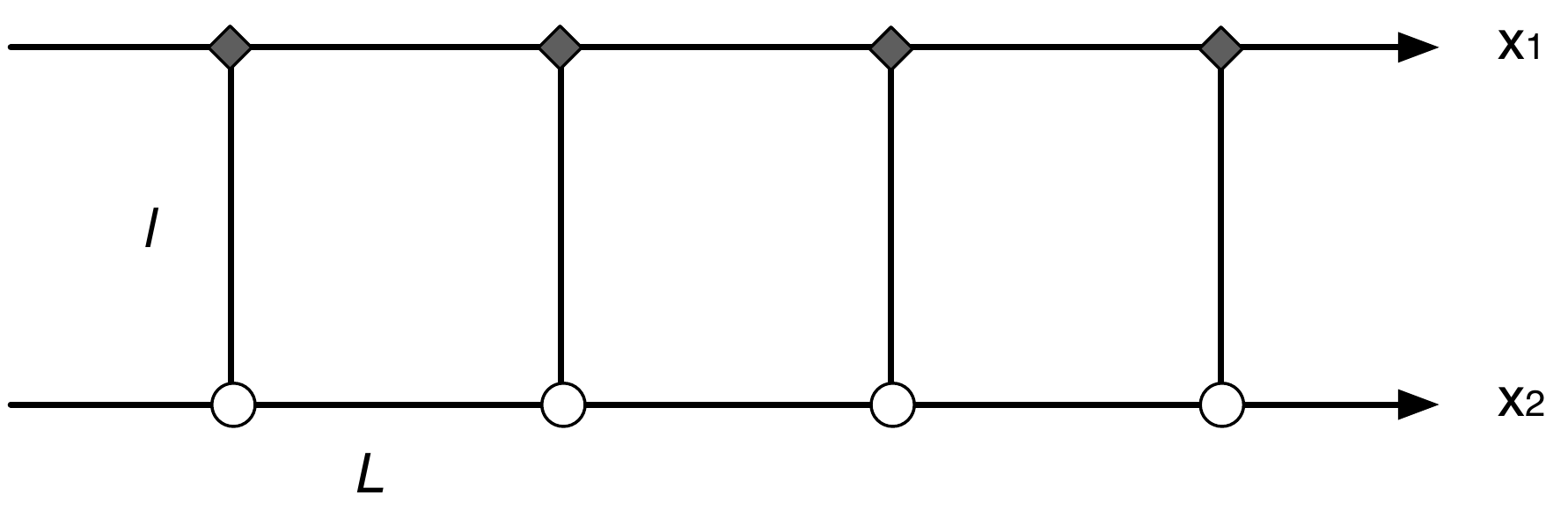} \\
\caption{Asymmetric quantum ladder with different kind of vertex couplings at black and white vertices}
\label{fig5}
\end{figure}

Consider a ladder-like quantum graph made up of two parallel lines linked at regular intervals by internal lines on both parallel lines (Fig. \ref{fig5}).
Applying the Bloch theorem again, 
we examine a single unit segment of the lattice, which is
described by coordinates $x_1 \in [0, L)$, $x_2 \in [0, L)$ and $x_\ell \in [0,\ell] $.  The wave function is given in the form
\begin{eqnarray}
\label{solwv11}
&
\psi_1(x_1) = \alpha_1 e^{i k x_1} + \beta_1 e^{-ikx_1} ,
\nonumber \\
&
\psi_2(x_2) = \alpha_2 e^{i k x_2} + \beta_2 e^{-ikx_2} ,
\nonumber \\
& 
\phi(x_\ell) = \alpha_\ell e^{i k x_\ell} + \beta_\ell e^{-ikx_\ell} .
\end{eqnarray}
In the unit segment there are two junctions, $(x_1=0$ $\&$ $x_\ell = 0)$ and $(x_2 = 0$ $\&$ $x_\ell = \ell)$.  At each junction we consider the vectors $\Psi_t$, $\Psi'_t$ and $\Psi_b$, $\Psi'_b$, which are respectively defined by
\begin{eqnarray}
\Psi_t = \begin{pmatrix} \psi_1(0_-) \\ \psi_1(0_+) \\ \phi(0) \end{pmatrix} \ ,
\Psi'_t = \begin{pmatrix} -\psi'_1(0_-) \\ \psi'_1(0_+) \\ \phi'(0) \end{pmatrix} \ ,
\label{psit}
\end{eqnarray}
and
\begin{eqnarray}
\Psi_b = \begin{pmatrix} \psi_2(0_-) \\ \psi_2(0_+) \\ \phi(\ell) \end{pmatrix} \ ,
\Psi'_b  = \begin{pmatrix} -\psi'_2(0_-) \\ \psi'_2(0_+) \\ -\phi'(\ell) \end{pmatrix} \ .
\label{psib}
\end{eqnarray}
The properties of the junctions are determined by the linear matrix equations of $\Psi$ and $\Psi'$. 
We consider different types of vertex interactions at top and bottom nodes of the ladder,
\begin{eqnarray}
\label{ladder_gen_int}
&\begin{pmatrix} 1 & 0 & a_1 \\ 0 & 1 & a_2 \\ 0 & 0 & 0 \end{pmatrix} 
\Psi'_t
-
\begin{pmatrix}  v_{11} & v_{12} & 0\\ v_{21} & v_{22} & 0\\ -a_1^* & -a_2^* & 1 \end{pmatrix} 
\Psi_t
= 0,&
\nonumber\\
&\begin{pmatrix} 1 & b_1 & b_2 \\ 0 & 0 & 0 \\ 0 & 0 & 0 \end{pmatrix} 
\Psi'_b
-
\begin{pmatrix}  u & 0 & 0\\ -b_1^* & 1 & 0\\ -b_2^* & 0 &1 \end{pmatrix} 
\Psi_b
= 0,&
\end{eqnarray}
where we used notations (\ref{psit}) (\ref{psib}) and had conditions $v_{11}=v_{11}^*,v_{22}=v_{22}^*,v_{12}=v_{21}^*,u=u^*$.

Let us consider the limit case $\ell << \frac{1}{k}$, where we have
\begin{eqnarray}
&
\phi(\ell) = \alpha_\ell e^{ik\ell}  + \beta_\ell e^{-ik\ell} \approx \alpha_\ell + \beta_\ell = \phi(0) ,
\nonumber \\
&
\phi'(\ell) = ik \alpha_\ell e^{ik\ell}  - ik \beta_\ell e^{-ik\ell} \approx ik \alpha_\ell -ik \beta_\ell = \phi'(0).
\end{eqnarray}
%
Expelling the function $\phi$, we rewrite matrix equations (\ref{ladder_gen_int}) in the form
\begin{eqnarray}
\label{ladder_gen_int_limit}
\begin{pmatrix} 
1/a_1 & -1/a_2 & 0 & 0 \\
1/a_1 & 0 & 1/b_2 & b_1/b_2 \\
0 & 0 & 0 & 0 \\
0 & 0 & 0 & 0
\end{pmatrix} 
\begin{pmatrix}
\psi_1' \\ \psi_2' \\ \psi_3' \\ \psi_4'
\end{pmatrix} 
-
\begin{pmatrix} \frac{v_{11}}{a_1}-\frac{v_{21}}{a_2} & \frac{v_{12}}{a_1}-\frac{v_{22}}{a_2} & 0 & 0 \\
\frac{v_{11}}{a_1} & \frac{v_{12}}{a_1} & \frac{u}{b_2} & 0 \\ 0 & 0 & -b_1^* & 1 \\ -a_1^* & -a_2^* & b_2^* & 0
\end{pmatrix} 
\begin{pmatrix}
\psi_1 \\ \psi_2 \\ \psi_3 \\ \psi_4
\end{pmatrix} 
= 0.
\end{eqnarray}
Applying transformations introduced in paper \cite{approx-gen}, the boundary conditions (\ref{ladder_gen_int_limit}) can be rewritten in the \emph{$ST$-form}

\begin{eqnarray}
\label{mod_bound}
\begin{pmatrix} 
1 & 0 & a_1/b_2 & b_1a_1/b_2  \\
0 & 1 & a_2/b_2 & b_1a_2/b_2  \\
0 & 0 & 0 & 0 \\
0 & 0 & 0 & 0
\end{pmatrix} 
\begin{pmatrix}
\psi_1' \\ \psi_2' \\ \psi_3' \\ \psi_4'
\end{pmatrix} 
-
\begin{pmatrix} v_{11}+\frac{u a_1 a_1^*}{b_2 b_2^*} & v_{12}+\frac{u a_1 a_2^*}{b_2 b_2^*} & 0 & 0 \\
 v_{21}+\frac{u a_1^* a_2}{b_2 b_2^*} &  v_{22}+\frac{u a_2 a_2^*}{b_2 b_2^*}&  0 & 0 \\
 -a_1^*/b_2^* & -a_2^*/b_2^* & 1 & 0 \\ -b_1^*a_1^*/b_2^* & -b_1^*a_1^*/b_2^* & 0 & 1
\end{pmatrix} 
\begin{pmatrix}
\psi_1 \\ \psi_2 \\ \psi_3 \\ \psi_4
\end{pmatrix} 
= 0.
\end{eqnarray}
Considering the special case $u=0,\, a_2=a_1,\,b_1=\pm 1, b_2=1$ and $v_{11}=v_{22}=\pm v_{12}$ we obtain the boundary conditions investigated in previous section.
Particularly, we can derive effective connection conditions with matrix coefficients (\ref{ST}), (\ref{cc-ddp}), if we insert vertex couplings at black and white nodes in the following way:
%
\begin{eqnarray}
\label{ladder_gen_int2}
&\begin{pmatrix} 1 & 0 & a \\ 0 & 1 & a \\ 0 & 0 & 0 \end{pmatrix} 
\Psi'_t
-
\begin{pmatrix}  c & cd & 0\\ cd & cd^2 & 0\\ -a & -a & 1 \end{pmatrix} 
\Psi_t
= 0,&
\nonumber\\
&\begin{pmatrix} 1 & b/a & 1 \\ 0 & 0 & 0 \\ 0 & 0 & 0 \end{pmatrix} 
\Psi'_b
-
\begin{pmatrix}  0 & 0 & 0\\ -b/a & 1 & 0\\ -1 & 0 &1 \end{pmatrix} 
\Psi_b
= 0.&
\end{eqnarray}

In general case, depending on the values of the parameters in connection conditions at graph vertices, it can be revealed a large variety of the spectrum of bound state and complicated flux behaviour through ladder lanes. Therefore, let us focus on an asymmetric ladder with more simplified boundary conditions at its vertices.
Particularly, we require $\delta$ potential-type connection conditions for the upper junctions:
\begin{eqnarray}
\label{conc1}
\begin{pmatrix} 1 & 1 & 1\\ 0 & 0 & 0\\ 0 & 0 & 0 \end{pmatrix} 
\Psi'_t
-
\begin{pmatrix}  v & 0 & 0\\ -1 & 1 & 0\\ -1 & 0 & 1 \end{pmatrix} 
\Psi_t
= 0 .
\end{eqnarray}
For the bottom junctions, we assume that the connection conditions are more ``exotic'', in order to see whether that produces any interesting phenomena.
Specifically, we impose the symmetrized $\delta'$-type condition
\begin{eqnarray}
\label{conc2}
\begin{pmatrix} 1 & 0 & -1 \\ 0 & 1 & -1 \\  0 & 0 & u  \end{pmatrix} 
\Psi'_b
-
\begin{pmatrix}  0 & 0 & 0\\ 0 & 0 & 0\\  1 & 1 & 1 \end{pmatrix} 
\Psi_b
= 0 .
\end{eqnarray}
\begin{figure}[ht]\center
\includegraphics[width=50mm]{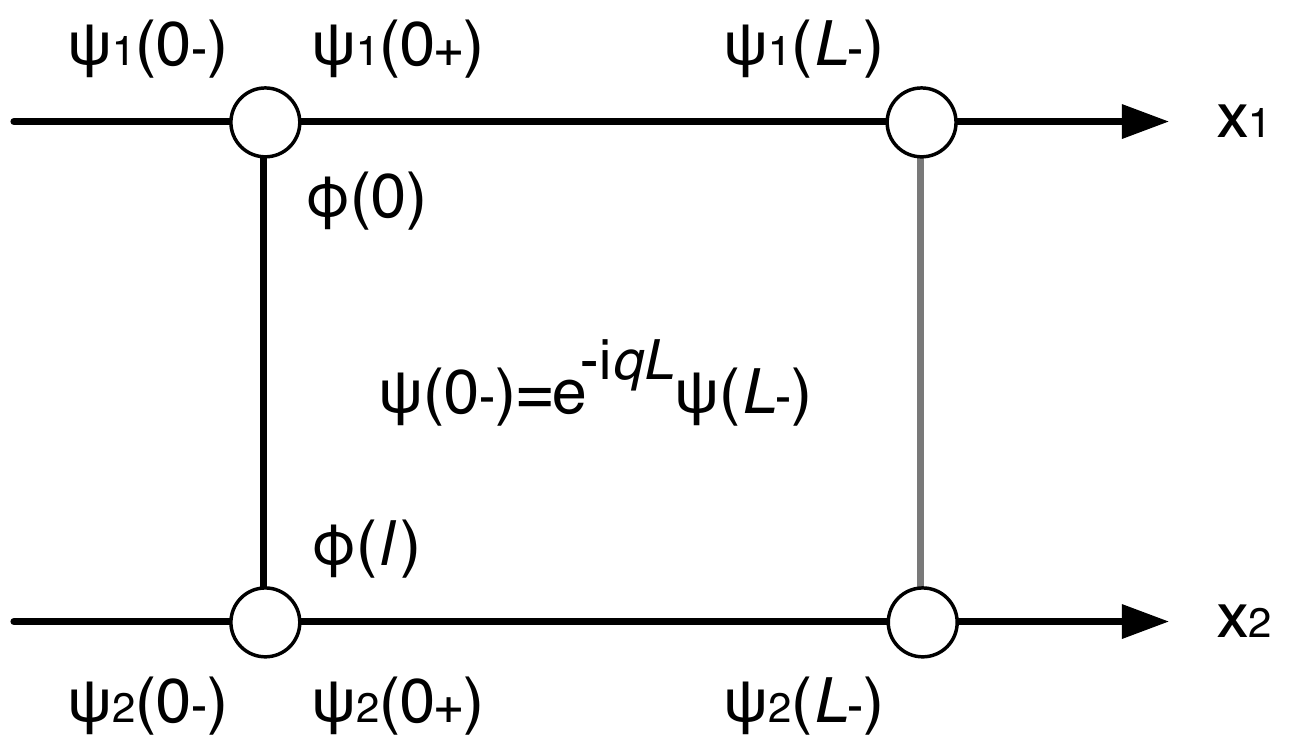} \\
\caption{Unit chain of the ladder lattice and the condition of quasi-periodicity of the wave function.}
\label{fig6}
\end{figure}
Now the problem became finding the solution (\ref{solwv1}) which would satisfy the connection conditions (\ref{conc1}),(\ref{conc2})
and quasi-periodicity of wave functions $\psi_1$ and $\psi_2$ (Fig \ref{fig6}). 
Substituting the limits of wave function (\ref{limfunc}) into the boundary conditions and completing a straightforward calculation, we derive
a relation between the energy $E = k^2$ and quasimomentum $q$
\begin{eqnarray}
\label{sec_ladder}
2 \cos (2 L q)-2 \cos (L q)\sin (k L) \left(k u+\frac{v}{k}\right) + \sin (2 k L) \left(k u-\frac{v}{k}\right)- \sin ^2 (k L)(u v+5)-2+\nonumber \\
+ \tan (k \ell ) \left(\sin ^2(k L) \left(\frac{v}{k}-k u\right)+2 \sin (2 k L)\right)=0.
\end{eqnarray}
From the secular equation (\ref{sec_ladder}), one can easily find the Bloch wave number $q$ by solving the quadratic equation.
For the given energy, we have either no real solution, one solution, or two solutions. Two solutions of secular equation are depicted in Fig. \ref{ladder_sol}(a)
by blue and red lines which have overlapping areas. Thus, we derived a band structure of spectrum containing gaps and zones with single or double solutions.


\begin{figure}[ht]\centering
\begin{tabular}{cc}
$\vcenter{\hbox{\includegraphics[width=65mm]{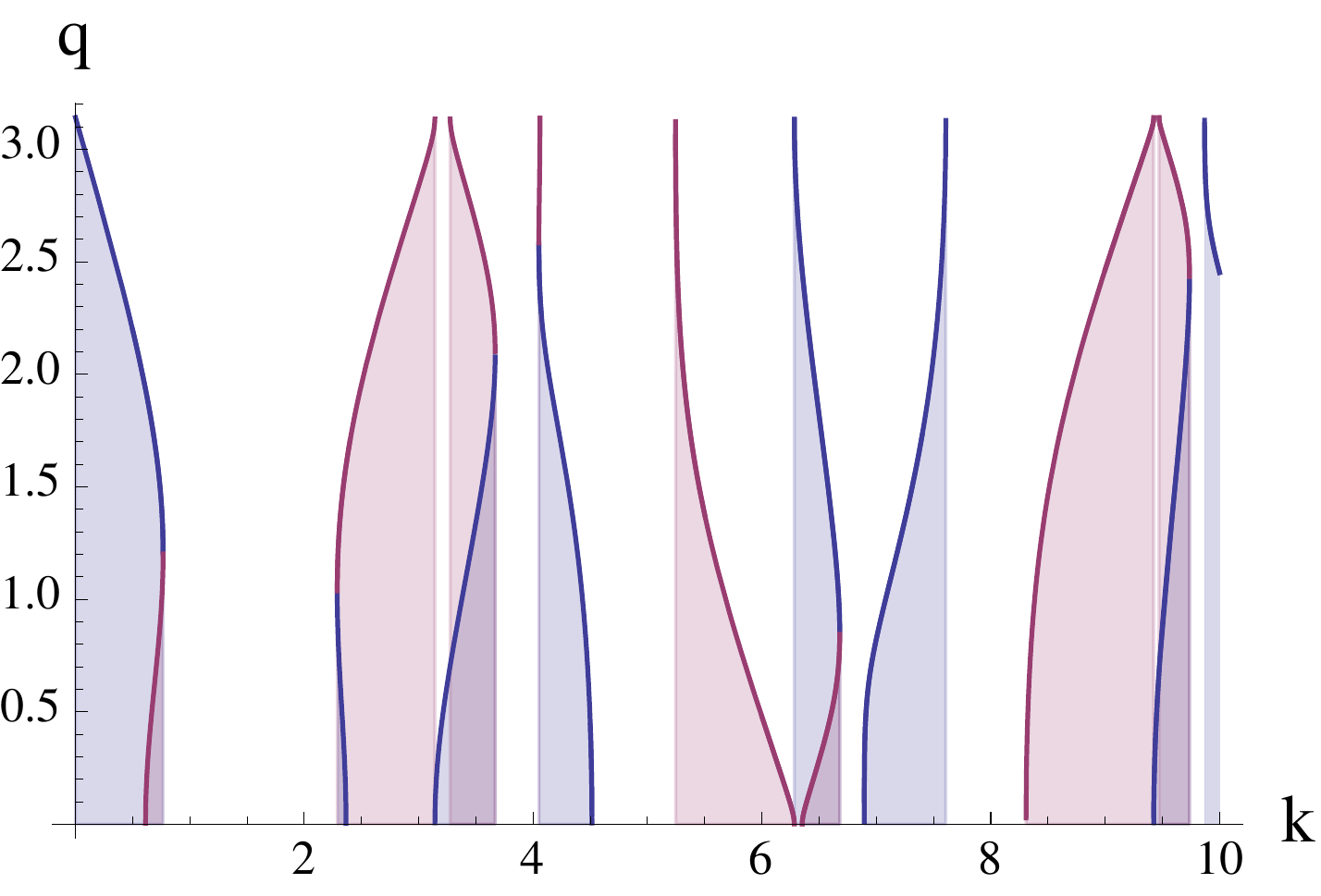}}}$
&
 $\vcenter{\hbox{\includegraphics[width=65mm]{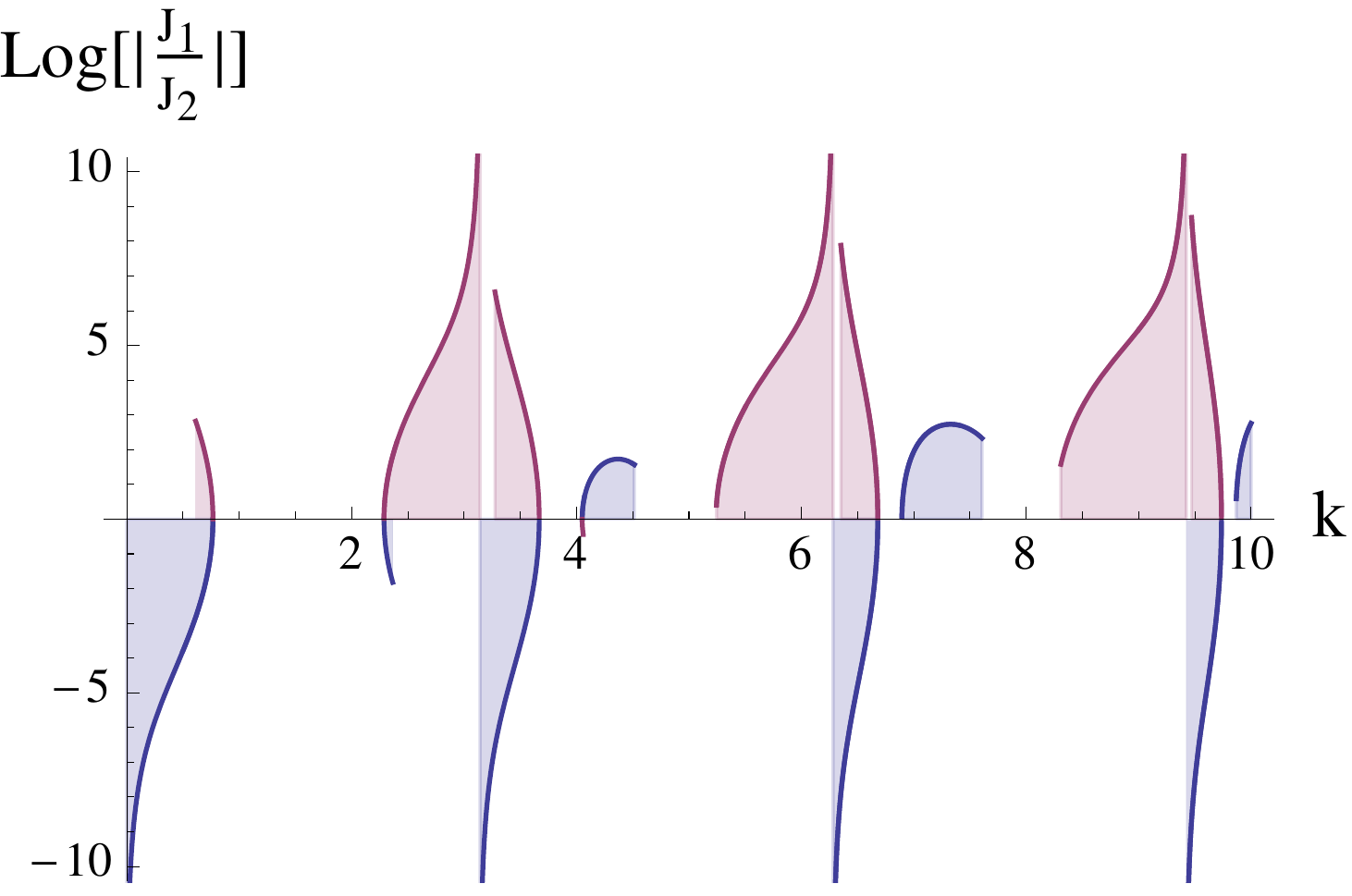}}}$ \\
(a)
&
(b)
\end{tabular}
\caption{Numerical plot of two solutions of secular equation shown by shaded red and blue lines (a) and the corresponding pattern of flux asymmetry (b)
for the following values of parameters: $u=1, v=1, L=1, \ell=1$.}
\label{ladder_sol}
\end{figure}
%

%
%
%
We calculate the quantum currents through upper and lower strands, $J_1$ and $J_2$ with  (\ref{currents12e}) just like in the previous section.
It turns out that the currents are always opposite to each other.  The dependence of the logarithm of the absolute value of the ratio $J_1$ and $J_2$ on
wave number $k$ is shown in Fig. \ref{ladder_sol}(b), which shows the flux asymmetry between two strands.
Thus, we conclude that the introduction of different types of couplings at vertices of lattice strands leads to alternating asymmetry of currents,
as we move from one band to another. Such flux behaviour indicates the importance of the role of boundary conditions at lattice vertices and
reminds one of the case of the chained ring model in the previous section.

%

\section{2D plane with $\delta$ and $\delta'$ layers}

In this section we will consider a Kronig-Penney model on a 2D rectangular lattice with lattice constants $L$ and $\ell$. We will assume an alternating layer structure with $\delta$ and $\delta'$
vertices. Here also, due to the Bloch theorem, it is sufficient to examine the wave function in the unit segment and use boundary conditions at $\delta$ and $\delta'$ connections.
Note that the lattice periods along $x$ and $y$ directions are $L$ and $2\ell$ respectively
 \begin{figure}[ht]\center
\includegraphics[width=50mm]{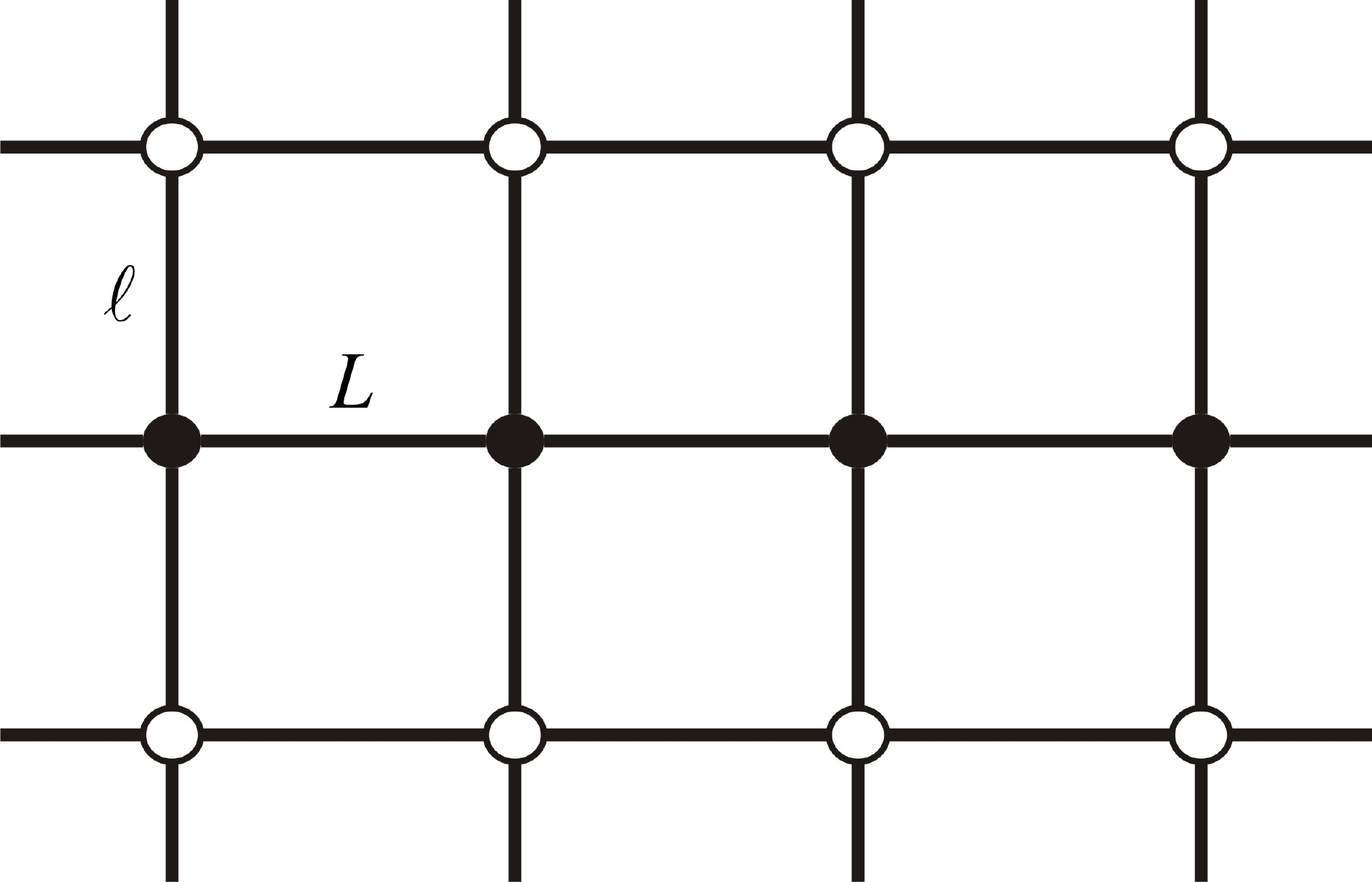}
\caption{2D extension of asymmetric quantum ladder with $\delta$ and $\delta'$ nodes.}
\label{fig8}
\end{figure}

\begin{eqnarray}
&\psi_1(x_1)=\alpha_1 e^{ikx_1}+\beta_1 e^{-ikx_1},&0<x_1<L,\nonumber\\
&\psi_2(x_2)=\alpha_2 e^{ikx_2}+\beta_2 e^{-ikx_2},&0<x_2<L,\nonumber\\
&\phi(y)=\alpha_3 e^{iky}+\beta_3 e^{-iky},& 0<y<\ell, \nonumber\\
&\phi(y)=\alpha_4 e^{iky}+\beta_4 e^{-iky},& \ell<y<2\ell.
 \end{eqnarray}
 As in previous sections, we use notations $\Psi_t$, $\Psi'_t$, $\Psi_b$, and $\Psi'_b$ defined by
\begin{eqnarray}
\Psi_t = \begin{pmatrix} \psi_1(0_-) \\ \phi(0_-) \\ \psi_1(0_+) \\ \phi(0_+) \end{pmatrix} \ ,
\Psi'_t = \begin{pmatrix} -\psi'_1(0_-) \\ -\phi'(0_-) \\ \psi'_1(0_+) \\ \phi'(0_+) \end{pmatrix} \ ,
\Psi_b = \begin{pmatrix} \psi_2(0_-)\\ \phi(\ell_-) \\ \psi_2(0_+) \\ \phi(\ell_+) \end{pmatrix} \ ,
\Psi'_b  = \begin{pmatrix} -\psi'_2(0_-)\\ -\phi'(\ell_-) \\ \psi'_2(0_+) \\ \phi'(\ell_+) \end{pmatrix} \ 
\end{eqnarray}
and write the boundary conditions for $\delta$ junction in the form
 \begin{eqnarray}
\label{2dcond1}
\begin{pmatrix} 1 & 1 & 1 & 1\\ 0 & 0 & 0 & 0\\ 0 & 0 & 0 & 0\\ 0 & 0 & 0 & 0 \end{pmatrix} 
\Psi'_t
-
\begin{pmatrix}  v & 0 & 0 & 0\\ -1 & 1 & 0 & 0\\ -1 & 0 & 1 & 0 \\ -1 & 0 & 0 & 1 \end{pmatrix} 
\Psi_t
= 0
\end{eqnarray}
and for the symmetrized $\delta'$-coupling
\begin{eqnarray}
\label{conc2}
\begin{pmatrix} 1 & 0 & 0 & -1 \\ 0 & 1 & 0 & -1 \\ 0 & 0 & 1 & -1\\ 0 & 0 & 0 & u  \end{pmatrix} 
\Psi'_b
-
\begin{pmatrix}  0 & 0 & 0 & 0\\ 0 & 0 & 0 & 0\\ 0 & 0 & 0 & 0 \\ 1 & 1 & 1 & 1 \end{pmatrix} 
\Psi_b
= 0 .
\end{eqnarray}
The condition of quasiperiodicity of wave function along $x$ and $y$ directions gives
\begin{eqnarray}
&
\psi_1(0_-) = e^{-i q_x L}( e^{i k L}\alpha_1 + e^{-i k L}\beta_1) ,\quad
\psi'_1(0_-) = ik e^{-i q_x L}( e^{i k L}\alpha_1 - e^{-i k L}\beta_1) ,
\nonumber \\
&
\psi_2(0_-) = e^{-i q_x L}( e^{i k L}\alpha_2 + e^{-i k L} \beta_2) ,\quad
\psi'_2(0_-) = ik e^{-i q_x L}( e^{i k L}\alpha_2 - e^{-i k L} \beta_2) ,
\nonumber \\
&
\phi(0_-) = e^{-2i q_y \ell}( e^{2i k \ell}\alpha_4 + e^{-2i k \ell}\beta_4) ,\quad
\phi'(0_-) = ik e^{-2i q_y \ell}( e^{2i k \ell}\alpha_4 - e^{-2i k \ell}\beta_4) ,
\label{2Dlimfunc}
\end{eqnarray}
To obtain the spectral equation, more tedious calculations are needed, compared to that of the ladder case.
However, the procedure is quite similar to previous case and the secular equation reads
\begin{eqnarray}
\cos (2 \ell q_y)=\frac{1}{16 \sin ^2(kL) } \Bigg( \sin (k L) \left(8 k u+\frac{8 v}{k}\right) (\cos (2 k \ell )+1) \cos \left(L q_x\right) \qquad\qquad\qquad\qquad\ \ 
\nonumber\\
- 8(\cos (2 k \ell )+1) \sin \left(2 L q_x\right)
+\left(\frac{4 v}{k}-4 k u\right) \Big(\sin (2 k L)-\sin (2 k \ell )+\sin (2 k (L+\ell ))\Big)\ 
\nonumber\\
+u v \cos (2 k (L-\ell ))-2 u v 
+(u v\!+\!16) \cos (2 k (L+\ell ))+(2 u v\!+\!8) \Big(\cos (2 k L)-\cos (2 k \ell )\Big)\Bigg).
\label{2DsecLl}
\end{eqnarray}
The expression (\ref{2DsecLl}) can be simplified if we assume $\ell=L$, which reduces it into
\begin{eqnarray}
\cos \left(2 L q_y\right)=\frac{1}{4 \sin^2(kL)} \Bigg(\left(\sin (k L)+\sin (3 k L)\right) \left(k u+\frac{v}{k}\right) \cos \left(L q_x\right)
\qquad\qquad\qquad\qquad
\nonumber\\
-2 (\cos (2 k L)+1) \cos \left(2 L q_x\right)+
\sin (4 k L) \left(\frac{v}{k}-k u\right)+\left(\frac{u v}{4}+4\right) \cos (4 k L)-\frac{u v}{4}\Bigg).
\label{2DsecLL}
\end{eqnarray}
Thus, we obtained a spectral equation, which gives a connection among Bloch wave numbers  $q_x$ and $q_y$, and the energy $E=k^2$. To define the state of the system, we need to fix not
only energy, but also one of the Bloch numbers; using the spectral equation (\ref{2DsecLl}), we can obtain the second Bloch number. Numerical calculations show that energy gaps appear in the spectrum, i.e. there is some range of values for $k$ for which the equation (\ref{2DsecLL}) has no real solution for $q_x$ and $q_y$.
All gaps have period $2\pi L$ and in the particular case $u=1, v=1, L=1$ we have $4.431\lesssim k \lesssim4.712$ for the first gap.

To compare particle flux on $x$ and $y$ directions, we have to take into account four types of currents 
\begin{eqnarray}
J_1^x=\frac{k\hbar}{m}\left(|\alpha_1|^2-|\beta_1|^2\right),\nonumber\\
J_2^x=\frac{k\hbar}{m}\left(|\alpha_2|^2-|\beta_2|^2\right),\nonumber\\
J_1^y=\frac{k\hbar}{m}\left(|\alpha_3|^2-|\beta_3|^2\right),\nonumber\\
J_2^y=\frac{k\hbar}{m}\left(|\alpha_4|^2-|\beta_4|^2\right).
\label{2Dcurrents12}
\end{eqnarray}
In the case of our model for the certain values of energies and Bloch wave numbers, we have no currents on either $x$ or $y$ directions.
For example, assuming again $u=1, v=1, \ell=L=1$ we found out $J_1^x=J_2^x=J_1^y=J_2^y=0$ for $q_x=1, k=\pi/2$. There are also
special values of energy and Bloch wave $q_x$, for which the particle current disappears only along the vertical or horizontal lattice bonds.  
For instance, $J_1^x=J_2^x=0, J_1^y\neq 0, J_2^y\neq 0$ when $q_x=\pi p\; (p=0,\pm 1, \pm 2,...)$ and $J_1^y=J_2^y=J_1^x=0, J_2^x\neq 0$ if $q_x=1, k\approx 0.753$
(the case $J_2^x=0, J_1^x\neq 0$ corresponds to $q_x=-1$).

\bigskip

\section{Realization of quantum graphs with $\delta$ vertices}

In this section we give a construction scheme for the physical realization of non-conventional vertex couplings of a double-stranded lattice. We will use an approximation method of arbitrary
singular vertices by quantum graphs carrying only $\delta$-couplings and constant vector potentials, \cite{approx-gen}
which is a generalisation of the ``singular local potential realisation'' of the $\delta'$ coupling on a line \cite{delta-prime, ENZ01}.  Referring the reader to the original papers for full details, let us very briefly review the basic points of the scheme:
\begin{itemize}
\item Separate $n$ edges of the star graph and connect nodes $i$ and $j$ by lines of the length $2d_0$ if one of the following three
conditions is satisfied
\begin{itemize}
\item[(1)] $j\leq m$, $k\geq m+1$, and $T_{jk}\neq0$
(or $j\geq m+1$, $k\leq m$, and $T_{kj}\neq0$),
\item[(2)] $j\leq m, k\leq m$ and
$(\exists l\geq m+1)(T_{jl}\neq0\wedge T_{kl}\neq0)$,
\item[(3)] $j\leq m, k\leq m$, $S_{jk}\neq0$, and the previous
condition is not satisfied,
\end{itemize}
where the rows of the matrix $T$ are indexed from 1 to $m$ and the columns from $m+1$ to $n$. 
\item
Apply a $\delta$-coupling with interaction parameter $v_j$ at the $j$-th node for all $1\leq j\leq n$ and add a $\delta$-potential with parameter $w_{jk}$
at the centre of the constructed line $(j,k)$ for all $1\leq k\leq N_j$, where $N_j$ is a number of vertices
joined to the $j$-th one. Place also constant vector potentials $A_{(j,l)}$ and $A_{(l,k)}=-A_{(j,l)}$ on the
segments $(j,l)$ and $(l,k)$ respectively, where $l$ is the centre of the line $(j,k)$ $1\leq l\leq N_j$.
\item
The values of $\delta$-interaction parameters $v_j$, $w_{jk}$ and vector potentials $A_{(j,k)}$ depend on size parameter $d_0$ and are determined
by the following way:
\begin{itemize}
\item[(1)]
If $j\leq m$ and $k\geq m+1$ then
\begin{eqnarray}\label{Avw1}
A_{(j,k)}(d_0)&=&\left\{
\begin{array}{lcl} \frac{1}{2d_0}\arg\,T_{jk} & \text{if} & Re\, T_{jk}\geq0,\nonumber \\
[.5em] \frac{1}{2d_0}\left(\arg\,T_{jk}-\pi\right) & \text{if} & Re\, T_{jk}<0,
\end{array}\right. \nonumber\\
v_k(d_0)&=&\frac{1-N_k+\sum_{h=1}^m\langle T_{hk}\rangle}{d_0}, \nonumber\\
w_{jk}(d_0)&=&\frac{1}{d_0}\left(-2+\frac{1}{\langle T_{jk}\rangle}\right),
\label{case1}
\end{eqnarray}
where the symbol $\langle\cdot\rangle$ means
\begin{equation*}
\langle c\rangle=\left\{\begin{array}{ccl}
|c| & \text{if} & Re\, c\geq0\,, \\
-|c| & \text{if} & Re\, c<0\,.
\end{array}\right.
\end{equation*}
\item[(2)]
In the case $j\leq m$, $k\leq m$ we have
\begin{eqnarray}\label{Avw2}
\!\!\!\!\!\!\!\!\!\!\!\!\!\!\!\!
A_{(j,k)}(d_0)\!\!\!\!&=&\!\!\!\!\left\{
\begin{array}{lcl} \frac{1}{2d_0}\arg\,\left(d_0 S_{jk}
+\sum_{l=m+1}^n T_{jl}\overline{T_{kl}}\right) \quad
\text{if}\  Re\left(d_0\cdot S_{jk}
+\sum_{l=m+1}^n T_{jl}\overline{T_{kl}}\right)\geq0\,,\nonumber\\
[.5em]
\frac{1}{2d_0}\left[\arg\,\left(d_0 S_{jk}
+\sum_{l=m+1}^n T_{jl}\overline{T_{kl}}\right)-\pi\right] \ 
\text{if}\  Re\left(d_0 S_{jk}
+\sum_{l=m+1}^n T_{jl}\overline{T_{kl}}\right)<0\,,
\end{array}\right. \nonumber\\
\!\!\!\!\!\!\!\!\!\!\!\!\!\!\!\!
v_j(d_0)\!\!\!\!&=&\!\!\!\!S_{jj}-\frac{N_j}{d_0}-\sum_{k=1}^m\left\langle
S_{jk}+\frac{1}{d_0}\sum_{l=m+1}^n
T_{jl}\overline{T_{kl}}\right\rangle
+\frac{1}{d_0}\sum_{l=m+1}^n(1+\langle T_{jl}\rangle)\langle
T_{jl}\rangle\,,\nonumber\\
\!\!\!\!\!\!\!\!\!\!\!\!\!\!\!\!
w_{jk}(d_0)\!\!\!\!&=&\!\!\!\!-\frac{1}{d_0}\left(2+\left\langle d_0\cdot S_{jk}
+\sum_{l=m+1}^n T_{jl}\overline{T_{kl}}\right\rangle^{-1}\right)\,.
\label{case2}
\end{eqnarray}
\end{itemize}
It can be shown that the constructed graph converges to desired singular vertex coupling in the norm-resolvent sense \cite{approx-gen}.
Also note that this approximation scheme and choice of parameters are not unique.
\end{itemize}
We represent singular vertices of double-stranded quantum graph in two alternative ways. For the sake of simplicity,
let us consider only the case $b=a$ and $d=1$ in the matrices (\ref{cc-ddp}).
At first, we directly apply the construction scheme (\ref{case1}), (\ref{case2}) to design a quantum star graph with connection conditions
\begin{eqnarray}
\label{ds-concondnew}
\begin{pmatrix}  c & c & 0 & 0 \\ 
 c & c & 0 & 0 \\
  -a & -a & 1 & 0 \\
 -a &  -a & 0 & 1 \end{pmatrix}
\Psi
-
\begin{pmatrix} 1 & 0 & a & a\\ 0 & 1 & a & a\\ 0 & 0 & 0 & 0\\ 0 & 0 & 0 & 0  \end{pmatrix} 
\Psi'
= 0
\end{eqnarray}
Then we give the second approximating way of the same vertices by implementing the asymmetric ladder system (see Section 3) with boundary conditions at the vertices
\begin{eqnarray}
\label{ladder_gen_int3}
&\begin{pmatrix} 1 & 0 & a \\ 0 & 1 & a \\ 0 & 0 & 0 \end{pmatrix} 
\Psi'_t
-
\begin{pmatrix}  c & c & 0\\ c & c & 0\\ -a & -a & 1 \end{pmatrix} 
\Psi_t
= 0,&
\nonumber\\
&\begin{pmatrix} 1 & 1 & 1 \\ 0 & 0 & 0 \\ 0 & 0 & 0 \end{pmatrix} 
\Psi'_b
-
\begin{pmatrix}  0 & 0 & 0\\ -1 & 1 & 0\\ -1 & 0 &1 \end{pmatrix} 
\Psi_b
= 0.&
\end{eqnarray}

\begin{figure}[ht]\centering
\begin{tabular}{cc}
$\vcenter{\hbox{\includegraphics[width=50mm]{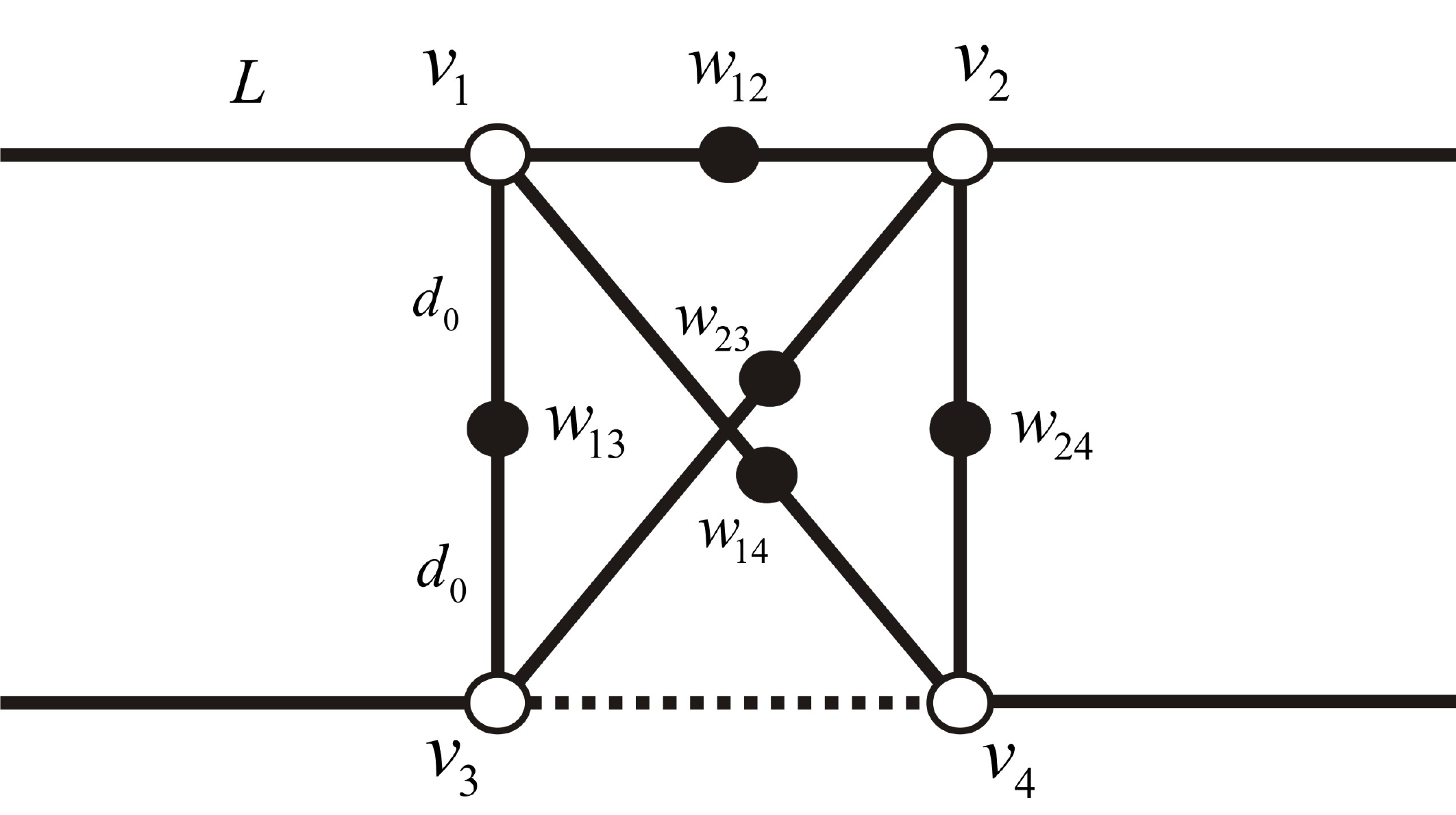}}}$
\hspace*{.5in}
&
 $\vcenter{\hbox{\includegraphics[width=35mm]{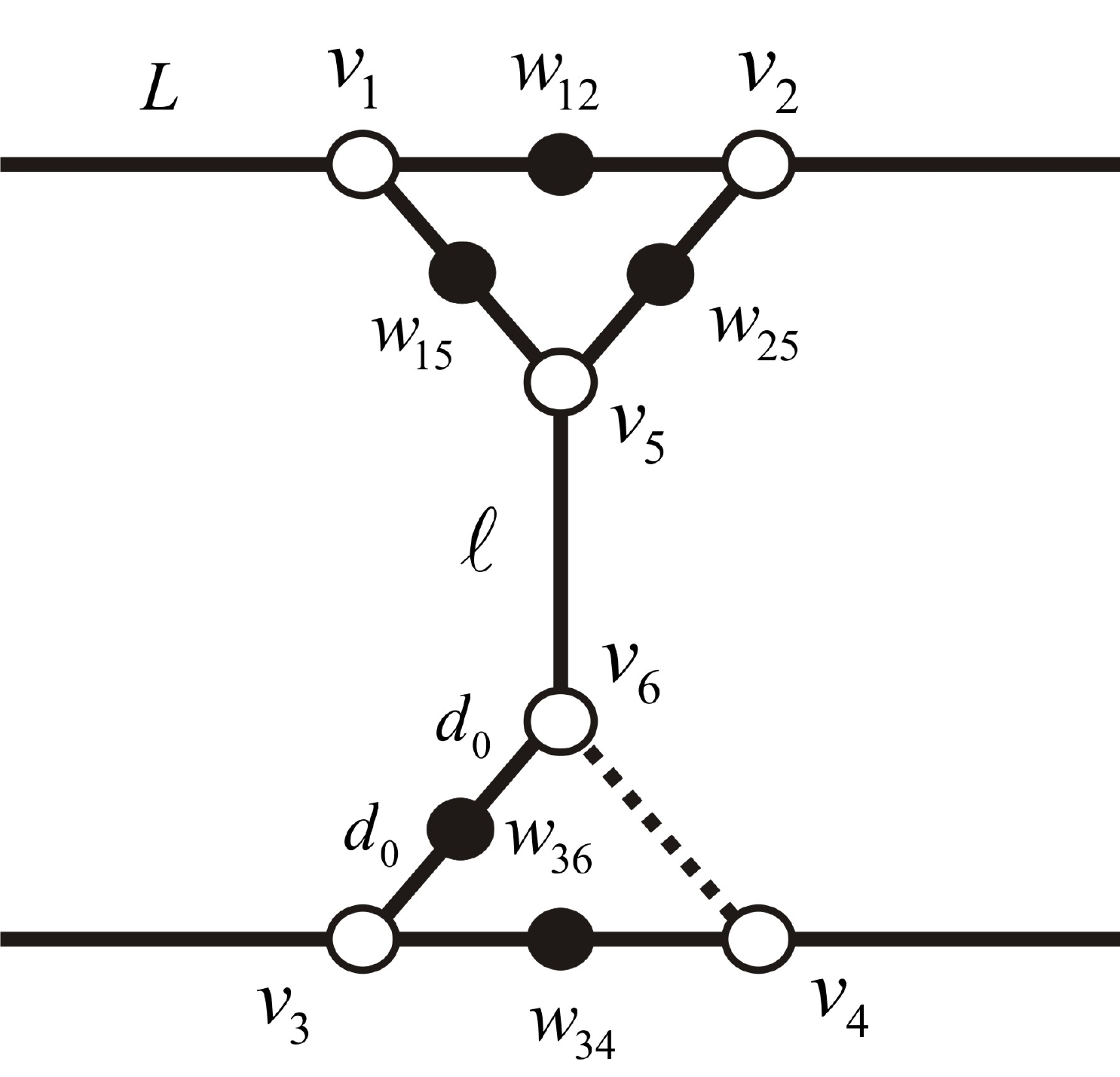}}}$ \\
(a)
\hspace*{.5in}
&
(b)
\end{tabular}
\caption{Two alternative ways of approximating a double-stranded junction by delta vertices.}
\label{delta}
\end{figure}

Let us split the graph nodes of the double-stranded chain and quantum ladder, as shown in figures \ref{delta}a and \ref{delta}b, and connect the free endpoints by wires.
Note that the edges shown by a dashed lines are not connected according to our construction scheme.
As all the parameters in boundary conditions (\ref{ds-concondnew}) and 
(\ref{ladder_gen_int3}) are real, we will not need to apply vector potentials to the connecting lines.
The values of $\delta$-potential strengths at nodes and in the centres of graph edges, are determined by
\begin{eqnarray}
\label{ds-potentials}
&&v_1=v_2=-c-\frac{1}{d_0}(2a^2-2a+3),\;\;\;\;\;\;\; v_3=v_4=\frac{1}{d_0}(-1+2a),\nonumber\\
&&w_{12}=-\frac{1}{d_0}(2+\frac{1}{d_0 c+2a^2}),\;\;\;\;\;\; w_{13}=w_{14}=w_{23}=w_{24}=\frac{1}{d_0}(-2+\frac{1}{a})
\end{eqnarray}
for the figure \ref{delta}a and
\begin{eqnarray}
\label{ds-potentials}
&&v_1=v_2=-c-\frac{1}{d_0}(a^2-a+2),\;\;\;\;\;v_3=0,\;\;\;\;v_4=v_6=\frac{1}{d_0},\;\;\;\; v_5=\frac{1}{d_0}(-1+2a), \nonumber\\
&&w_{12}=-\frac{1}{d_0}(2+\frac{1}{d_0 c+a^2}),\;\;\;\;\; w_{15}=w_{25}=\frac{1}{d_0}(-2+\frac{1}{c}),\;\;\;\;\;\; w_{34}=w_{36}=-\frac{1}{d_0}
\end{eqnarray}
for the figure \ref{delta}b. Thus, the boundary conditions (\ref{ds-concondnew}) can be achieved in the shrinking limit $d_0<<\frac{1}{k}$ for the double-stranded chain and $d_0<<\ell<<\frac{1}{k}$ for the ladder lattice.  

\section{Summary and prospects}

In this paper, we have considered minimal double-stranded extensions of the celebrated Kronig-Penney model and uncovered unexpected exotic features in the pattern of fluxes that go through the two strands.  The key to the discovery has been the employment of non-conventional vertex couplings, which, at the first glance, seems rather artificial.  We have shown, however, that the model can be regarded as a mathematical abstraction of a more realistic periodic two-lane ladder with conventional $\delta$-couplings, albeit with more complicated internal graph structures.

The investigation of straightforward extensions to the chains with triple and more strands can be accomplished by direct computations.  The more challenging problem is the full analysis of a layered two-dimensional Kronig-Penney lattice.  The identification of parameter ranges which permit a symmetric flux in two dimensional models should be considered of great importance for designing and developing quantum devices.

\section*{Acknowledgments}
%
The authors thank Michelle Noyes for technical assistance in the preparation of this paper.
This research was supported by the Japan Ministry of Education, Culture, Sports, Science and Technology under the Grant number 24540412.

%

%

\end{document}